\documentstyle[12pt,axodraw]{article}
\newcommand{\NP}{Nucl. Phys. }
\newcommand{\PR}{Phys. Rev. }
\newcommand{\PRL}{Phys. Rev. Lett. }
\newcommand{\PL}{Phys. Lett. }

\begin{document}
\baselineskip=20pt

\pagenumbering{arabic}

\vspace{1.0cm}
\begin{center}
{\Large\sf Pair production of neutral Higgs bosons through
noncommutative QED interactions at linear colliders}\\[10pt]
\vspace{.5 cm}

{Harald Grosse}
\vspace{1.0ex}

{\small Institut f\"ur Theoretische Physik, Universit\"at Wien,\\
Boltzmanngasse 5, A-1090 Wien, Austria\\}

\vspace{3.0ex}
{Yi Liao}
\vspace{1.0ex}

{\small Institut f\"ur Theoretische Physik, Universit\"at Leipzig,
\\
Augustusplatz 10/11, D-04109 Leipzig, Germany\\}

\vspace{2.0ex}

{\bf Abstract}
\end{center}

We study the feasibility of detecting noncommutative (NC) QED
through neutral Higgs boson ($H$) pair production at linear 
colliders (LC). This is based on the assumption that $H$ 
interacts directly with photon in NCQED as suggested by symmetry
considerations and strongly hinted by our previous study on 
$\pi^0$-photon interactions. We find the following striking 
features as compared to the standard model (SM) result:
(1) generally larger cross sections for an NC scale of order $1$ 
TeV; 
(2) completely different dependence on initial beam polarizations;
(3) distinct distributions in the polar and azimuthal angles; and
(4) day-night asymmetry due to the Earth's rotation.
These will help to separate NC signals from those in the
SM or other new physics at LC. We emphasize the importance of 
treating properly the Lorentz noninvariance problem and show how 
the impact of the Earth's rotation can be used as an advantage
for our purpose of searching for NC signals.

\begin{flushleft}


\end{flushleft}

\newpage

Noncommutative (NC) field theories have recently received a lot of 
attention mainly because of their connection to string theories 
$\cite{string}$, but they are certainly interesting in their own 
right. A possible way to construct the NC version of a field theory 
from its ordinary commutative counterpart is by replacing the usual 
product of fields in the action with the $\star$-product of fields. 
The $\star$-product of the two fields $\phi_1(x)$ and $\phi_2(x)$ 
is defined as 
\begin{equation}
\displaystyle(\phi_1\star\phi_2)(x)=\left[\exp\left(i/2~
\theta^{\mu\nu}\partial^x_{\mu}\partial^y_{\nu}\right)
\phi_1(x)\phi_2(y)\right]_{y=x},
\end{equation}
where $\theta^{\mu\nu}$ is a real antisymmetric constant matrix that
parametrizes the noncommutativity of spacetime,
\begin{equation}
\displaystyle[x^{\mu},x^{\nu}]=i\theta^{\mu\nu},
\end{equation}
and has dimensions of length squared.

NC quantum electrodynamics (NCQED) of photons and electrons is 
then given by the following Lagrangian $\cite{ncqed}$:
\begin{equation}
\begin{array}{rcl}
{\cal L}&=&\displaystyle-\frac{1}{4}F^{\mu\nu}\star F_{\mu\nu}+
\bar{\psi}\star(\gamma^{\mu}iD_{\mu}-m)\psi,\\
\end{array}
\end{equation}
with
$F_{\mu\nu}=\partial_{\mu}A_{\nu}-\partial_{\nu}A_{\mu}
+ie[A_{\mu},A_{\nu}]_{\star}$
and 
$D_{\mu}\psi=\partial_{\mu}\psi+ieA_{\mu}\star\psi$,
where the Moyal brackets are defined as
$[\phi_1,\phi_2]_{\star}=\phi_1\star\phi_2-\phi_2\star\phi_1$.
The action $\displaystyle\int d^4x~{\cal L}$ is invariant under 
the generalized $U(1)$ gauge transformation
\begin{equation}
\begin{array}{rcrcl}
\displaystyle
A_{\mu}&\to& A^{\prime}_{\mu}&=&U\star A_{\mu}\star U^{-1}+
ie^{-1}U\star\partial_{\mu}U^{-1},\\
\psi&\to& \psi^{\prime}&=&U\star\psi, 
\end{array}
\end{equation}
with 
$U(x)=(\exp[i\lambda(x)])_{\star}$, under which $F_{\mu\nu}$ 
also undergoes a nontrivial transformation,
$F_{\mu\nu}\to F^{\prime}_{\mu\nu}=U\star F_{\mu\nu}\star U^{-1}$.
Note that the neutral photon interacts with itself due to the
Moyal bracket term in $F_{\mu\nu}$ as in the usual non-Abelian 
gauge theory. 

NC field theories have rich phenomenological implications due to 
the appearance of new interactions and Lorentz noninvariance 
introduced by the constant $\theta_{\mu\nu}$ matrix. Some aspects 
of NCQED have been explored recently. From the point of view of 
effective field theories new physics effects amounts to 
introduction of some high dimension operators made up of 
ordinary fields and proportional to $\theta_{\mu\nu}$. Since the
latter carries two negative units in mass the effects are 
suppressed by two factors of some energy scale $\Lambda_{\rm NC}$
at which the noncommutativity sets in. At low energies they 
could only be detectable with precisely measured quantities, 
e.g., in some atomic systems $\cite{atomic}$. However, the 
suppression becomes less severe at high energy linear colliders 
(LC) if $\Lambda_{\rm NC}$ is not much larger than the collider's
energy. Considering the connection to string theories 
$\cite{string}$ and the possibility that gravitational and gauge
interactions may unite at a scale of order $1$ TeV in the 
framework of string theories $\cite{tev}$, it is reasonable to 
expect that the NC effects may also enter into the game at a 
similar scale. Along this line of argument some authors have 
discussed the feasibility of detecting NC signals at future LC 
through corrections to standard model 
(SM) processes $\cite{collider}$. Indeed
they found that with a collider energy of $0.5\sim 1.5$ TeV and 
an integrated luminosity of a few hundred fb$^{-1}$ it is 
possible to probe $\Lambda_{\rm NC}$ up to a few TeV at $95\%$ 
C.L. In this study we will work in the same spirit, but we will 
present some striking features of the process considered here 
which will be very helpful in distinguishing NC signals from 
those of the SM or other new physics.
We will emphasize the important issue of Lorentz noninvariance
in collider measurements and show how the impact of the Earth's 
rotation can be used for our purpose of detecting NC signals.

Another motivation derives from a recent work in which we showed 
how a simple and reasonable generalization of the anomalous 
$\pi^0$-photon interaction can lead to the three photon decay of 
the $\pi^0$ in NCQED $\cite{pion}$. The idea has got some 
support from analysis of anomalies in NC gauge theories 
$\cite{nca}\cite{nca2}$. We found that for the consideration to 
be physically self-consistent it is mandatory to treat the 
electrically neutral photon and $\pi^0$ on the same footing; 
namely the $\pi^0$ field must also undergo the same nontrivial 
transformation under $U(1)$ as the photon as if they were in the
adjoint representation of an effectively non-Abelian gauge 
theory,
\begin{equation}
\displaystyle
\phi^0\to \phi^{0\prime}=U\star\phi^0\star U^{-1},
\end{equation}
where $\phi^0$ stands for the $\pi^0$ field, so that neutral 
particles also participate in electromagnetic interactions,
\begin{equation}
\displaystyle
{\cal L}_{\phi^0}=\frac{1}{2}D_{\mu}\phi^0\star D^{\mu}\phi^0,
\end{equation}
with 
$D_{\mu}\phi^0=\partial_{\mu}\phi^0+ie[A_{\mu},\phi^0]_{\star}$.
This is reminiscent of the wisdom in the usual quantum field 
theory that one must keep all possible interactions that are 
consistent with symmetries for the theory to be renormalizable. 
Indeed it is far from clear at the moment how to extend the 
electroweak SM to NC spacetime consistently although there are 
already theoretical efforts $\cite{ncsm}$ and even 
phenomenological analysis on flavor physics in this direction 
$\cite{flavor}$. However we believe that the impressive lesson 
learnt from $\pi^0$ should be general enough to be applicable to
other neutral particles in the SM if it permits any kind of 
generalization to the NC case. Then the nice feature of 
uniformness and completeness among neutral particles concerning 
their electromagnetic interactions can be preserved in the 
generalized SM. This is especially true of the Higgs boson which
triggers the electroweak symmetry breakdown to the 
electromagnetic $U(1)$. Then Eqs. $(5)$ and $(6)$ apply equally 
well to the Higgs field $H$.

A direct result of interactions $(3)$ and $(6)$ with $\phi^0$ 
now denoting the neutral Higgs $H$ with mass $m_H$ is the 
occurence of the following process at the tree level:
\begin{equation}
e^-(k_1,\lambda_1)+e^+(k_2,\lambda_2)\to H(p_1)+H(p_2),
\end{equation}
where $k_i(p_i)$ are incoming (outgoing) momenta, and 
$\lambda_i=\pm 1$ are initial state helicities. 
The process proceeds through the $s$-channel exchange of photon,
whose amplitude is given by
\begin{equation}
\displaystyle{\cal A}_{\lambda_1\lambda_2}=
ie^{-(i/2)k_1\theta k_2}
\frac{4e^2}{s}\sin\left(\frac{1}{2}p_1\theta p_2\right)
\bar{v}\rlap/p_1P_2P_1u,
\end{equation}
where $u,~v$ are spinors, 
$P_{1,2}=(1\pm\lambda_{1,2}\gamma_5)/2$,
and $\sqrt{s}$ is the center-of-mass (c.m.) energy. We have used 
the abbreviation $p\theta q=\theta_{\mu\nu}p^{\mu}q^{\nu}$. Let 
us first work in the c.m. frame and denote as $(\theta,\phi)$ the 
polar and azimuthal angles of the Higgs boson. (This $\theta$ 
should not be confused with the NC parameter $\theta_{\mu\nu}$.)
Due to momentum conservation and antisymmetry of 
$\theta_{\mu\nu}$ only the components 
$\theta^{0i}\equiv(\vec{\theta})^i$ can contribute. 
Without loss of generality we assume that $\vec{\theta}$ lies in 
the $xz$ plane and deviates from the $z$ axis (i.e. the $e^-$ 
beam direction) by an angle $\gamma\in[0,\pi]$. For the 
parameters to be considered later, it is appropriate to use 
$\sin(p_1\theta p_2/2)\approx p_1\theta p_2/2$. 
The differential cross sections with polarized or unpolarized 
beams are
\begin{equation}
\begin{array}{rcl}
\displaystyle\left[\frac{d\sigma}{d\Omega}\right]_{\lambda_1\lambda_2}
&=&(1-\lambda_1\lambda_2)
\displaystyle\left[\frac{d\sigma}{d\Omega}\right]_{\rm unpol},\\
\displaystyle\left[\frac{d\sigma}{d\Omega}\right]_{\rm unpol}
&=&\displaystyle\frac{\alpha^2\beta^5}{64s}(s|\vec{\theta}|)^2
(s_{\gamma}s_{\theta}c_{\phi}+c_{\gamma}c_{\theta})^2
s^2_{\theta},
\end{array}
\end{equation}
where the factor $1/2!$ for identical particles has been 
included, $\beta=\sqrt{1-4m^2_H/s}$ is the Higgs boson velocity, 
and $c_{\theta}=\cos\theta$, $s_{\theta}=\sin\theta$, etc. The 
factors $\beta^3 s^2_{\theta}$ are due to the scalar nature of 
the Higgs boson and phase space while the additional factor 
$\beta^2$ and the one in brackets are peculiar to the NCQED 
interaction ${\cal L}_{\phi^0}$. The nontrivial azimuthal angle 
dependence arises because rotational invariance is broken by the
preferred direction of $\vec{\theta}$. The total cross sections 
are
\begin{equation}
\begin{array}{rcl}
\sigma_{\lambda_1\lambda_2}&=&\displaystyle(1-\lambda_1\lambda_2)
\sigma_{\rm unpol},\\
\sigma_{\rm unpol}&=&\displaystyle
\sigma_0\left(1-c^2_{\gamma}/2\right),\\
\sigma_0&=&\displaystyle\frac{\pi\alpha^2}{60}\beta^5 s 
|\vec{\theta}|^2.
\end{array}
\end{equation}
The proportionality to $s$ causes no problem with unitarity 
since it arises from an approximation which does not hold at 
very high energy.

One might expect to use Eqs. $(9)$ and $10$ for numerical 
analysis as was done in the literature $\cite{collider}$.
However, the above results are not directly applicable to a 
practical collider experiment. This is important because it 
would result in an incorrect interpretation of data.
Furthermore, as shown below this would also cause an 
unnecessary loss of information specific to NC signals.
Since $\theta_{\mu\nu}$ is not a 
Lorentz tensor and is given in some a priori frame, it should 
change from one frame to another differently from a tensor. For 
a practical collider experiment it takes a much longer time than
a day to collect data so we may expect important impacts from 
the Earth's rotation on data analysis. In the case considered 
here, the particles involved move much faster 
than the Earth's rotation, therefore we can safely ignore the 
change in magnitude of $\vec{\theta}$ in the local c.m. frame. But
we must take into account the change of its direction relative 
to the local frame, or, to put it more correctly, the rotation 
of our local frame as the Earth rotates.

Let us denote as $\rho\in[0,\pi]$ the angle between 
$\vec{\theta}$ and the Earth's rotation axis $\vec{R}$, which is
fixed to good precision. The location of the lab is described in
terms of two angles: the latitude $\sigma\in[-\pi/2,\pi/2]$ with
positive (negative) $\sigma$ denoting northern (southern) 
hemisphere, and the Earth's rotation angle (longitude)
$\omega\in[0,2\pi)$ measured relative to the plane spanned by
$\vec{R}$ and $\vec{\theta}$. Suppose the collider beam has an 
angle $\delta\in[0,2\pi)$ from the local longitudinal direction. 
The angles $\sigma$ and $\delta$ are fixed for a given 
collider. Then, we have
\begin{equation}
c_{\gamma}=-s_{\rho}(c_{\delta}s_{\omega}+
s_{\delta}s_{\sigma}c_{\omega})+c_{\rho}s_{\delta}c_{\sigma}.
\end{equation}

Upon considering the Earth's rotation we may have two types of
distributions, one in the local angles $\theta$ and $\phi$, and 
the other in $\omega$ of the Earth's rotation. For the former we
merely have to average over the rotation and find for the
unpolarized cross section,
\begin{equation}
\begin{array}{l}
\displaystyle\frac{4\pi}{\sigma_0}\overline{
\left[\frac{d\sigma}{d\Omega}\right]}=f(\theta,\phi),\\
f(\theta,\phi)=\displaystyle \frac{15}{4}
\left(\overline{s^2_{\gamma}}s^2_{\theta}c^2_{\phi}+
\overline{c^2_{\gamma}}c^2_{\theta}+
\overline{s_{\gamma}c_{\gamma}}s_{2\theta}c_{\phi}\right)
s^2_{\theta},
\end{array}
\end{equation}
where $s_{2\theta}=\sin(2\theta)$, etc. This amounts to analyzing
data as is usually done. 
To better describe the impact of the Earth's 
rotation we define the following day-night 
asymmetry as a periodic function of $\omega$ or time $t$,
\begin{equation}
\begin{array}{rcl}
A_{\rm DN}(\omega_a,\omega_b)&=&\displaystyle\frac
{\left[\int_{\omega_a}^{\omega_b}d\omega-
\int_{\omega_a+\pi}^{\omega_b+\pi}d\omega\right]\sigma(\omega)}
{\left[\int_{\omega_a}^{\omega_b}d\omega+
\int_{\omega_a+\pi}^{\omega_b+\pi}d\omega\right]\sigma(\omega)}\\
&=&\displaystyle\frac{N(\omega_b)-N(\omega_a)}{D(\omega_b)-D(\omega_a)},
\end{array}
\end{equation}
where $\sigma(\omega)$ is given in Eq. $(10)$ and
\begin{equation}
\begin{array}{rcl}
N(x)&=&(-c_{\delta}c_x+s_{\delta}s_{\sigma}s_x)
s_{2\rho}s_{\delta}c_{\sigma},\\
D(x)&=&-xc^2_{\rho}s^2_{\delta}c^2_{\sigma}+2x-
s^2_{\rho}/4\left[2x(s^2_{\delta}s^2_{\sigma}+c^2_{\delta})\right.\\
&&\left.+(s^2_{\delta}s^2_{\sigma}-c^2_{\delta})s_{2x}-
s_{2\delta}s_{\sigma}c_{2x}\right].
\end{array}
\end{equation}
And the integrated asymmetry is simply
\begin{equation}
A_{\rm DN}(0,\pi)=\displaystyle\frac{s_{2\rho}s_{2\delta}
c_{\sigma}}{2\pi\left\{1-\frac{1}{4}\left[s^2_{\rho}(c^2_{\delta}+s^2_
{\delta}s^2_{\sigma})+2c^2_{\rho}s^2_{\delta}c^2_{\sigma}\right]
\right\}}.
\end{equation}

Now let us examine the numerical significance of the above 
results. The scale of cross section is set by $\sigma_0$ which 
is plotted in Fig. $1$ as a function of $m_H$ at 
$\sqrt{s}=0.5,~1,~1.5$ TeV and for 
$\Lambda_{\rm NC}=|\vec{\theta}|^{-1/2}=1$ TeV. 
This should be contrasted with the SM result $\cite{eehh}$ which
is $0.1\sim 0.2$ fb for $m_H<2m_W$ at $\sqrt{s}=0.5$ TeV and for
the whole mass range shown at higher $\sqrt{s}$. We see that 
with $\sqrt{s}=1$ TeV or higher the NC signal dominates over the
SM background for an intermediate-mass Higgs boson. Since such 
processes will be searched for only after the Higgs boson has 
been found in its main production channels, the relatively low 
cross section can be compensated for by some knowledge of Higgs 
properties and by a high luminosity feasible at future LC. If 
the beams are properly polarized, the situation can even be 
better. Since NCQED interactions conserve helicity, we expect 
equal contributions from left-handed (LH) and right-handed (RH) 
polarized electron beams. For example, with RH electron and LH
positron beams we have a cross section twice as large as the 
unpolarized one [see Eq. $(10)$]. In the SM, the same process is
overwhelmingly dominated by one-loop $W^{\pm}$ boxes so that the
cross section for RH electron and LH positron beams is smaller 
by at least one order of magnitude than in the oppositely 
polarized case. Thus with suitably polarized beams one can earn
a signal over background ratio of a few tens even before a
cutoff is imposed. This already makes the process considered
here much more advantageous than those considered so far.

Some knowledge of Higgs properties beforehand also helps to 
reconstruct the final state of the process and to analyze the 
distribution of primary Higgs bosons. In Fig. $2(a)$ [$2(b)$] we 
plot the distribution $f(\theta,\phi)$ in the local angle 
$\theta$ ($\phi$) at a specified value of $\phi=\pi/4$ 
($\theta=\pi/4$) after averaging over the Earth's rotation. Note
that $f(\theta,\phi)$ depends only on orientation parameters.
We consider three sets of them for illustration:
$(1)$ $\rho=\pi/2$, $\delta=0$ and $\sigma$ free;
$(2)$ $\rho=0$, $\delta$ and $\sigma$ free but
$s_{\delta}c_{\sigma}=1/\sqrt{2}$; and 
$(3)$ $\rho=\delta=\sigma=\pi/4$.
Also shown is the distribution further averaged over $\phi$ 
($\theta$) for the parameter set $(3)$. In practice $\sigma$ 
and $\delta$ are known for a given collider; therefore we can 
fit the two distributions for just one angle $\rho$, which will 
balance the relative rareness of data in determining the 
relative direction of $\vec{\theta}$ to $\vec{R}$. 
The distribution can also
be easily discriminated from the SM one which follows 
approximately the $\sim\sin^2\theta$ law $\cite{eehh}$. Even if 
the NC signal accidentally shares a similar $\theta$ dependence 
(after averaging over $\phi$) with some other new physics signals, 
they can still be discriminated by $\phi$ dependence since the 
latter are trivial in $\phi$ dependence due to Lorentz invariance.

The above feature is further strengthened by the day-night 
asymmetry, shown in Fig. $3$ as histograms 
binned per half an hour for two sets of orientation parameters: 
the above case $(3)$, and 
$(4)$ $\rho=\pi/4$, $\delta=3\pi/4$ and $\sigma=0$.
Note that there is no asymmetry at $\rho=0,~\pi/2,~\pi$, or
$\delta=0,~\pi$, or $\sigma=\pm\pi/2$ and additionally 
at $\delta=\pi/2,~3\pi/2$ for the integrated asymmetry. For most
of these unfortunate orientations we can still
observe the periodic variation of the cross section with the 
Earth's rotation. Since this asymmetry or periodic variation
arises from Lorentz noninvariance in NCQED, it may be readily
separated from the null results in ordinary theories like the
SM and beyond.

The search for Higgs bosons and NC gauge theories are important
topics that attract a lot of attention. We attempted here to 
connect them through an analysis of neutral Higgs pair production 
at LC. The result is quite encouraging. We identified the salient
features of the process which proves to be much more advantageous
than others considered so far in that one can have a good 
signal-background (S/B) ratio with unpolarized beams and an even
excellent S/B with suitably polarized beams, and is thus unique 
in search for NC signals at LC. We have described for the first 
time how a practical measurement is affected by the Earth's 
rotation and how this impact may be used as an advantage in
discriminating NC signals from those in the ordinary commutative 
theories like SM and other new physics.

\vspace{0.5cm}
We are grateful to Klaus Sibold for many 
encouraging and helpful discussions and for carefully reading the
manuscript. H.G. enjoyed the stay at ITP, Universit\"at Leipzig 
where part of work was done.

\newpage
\begin{flushleft}
{\Large Figure Captions }
\end{flushleft}

\noindent
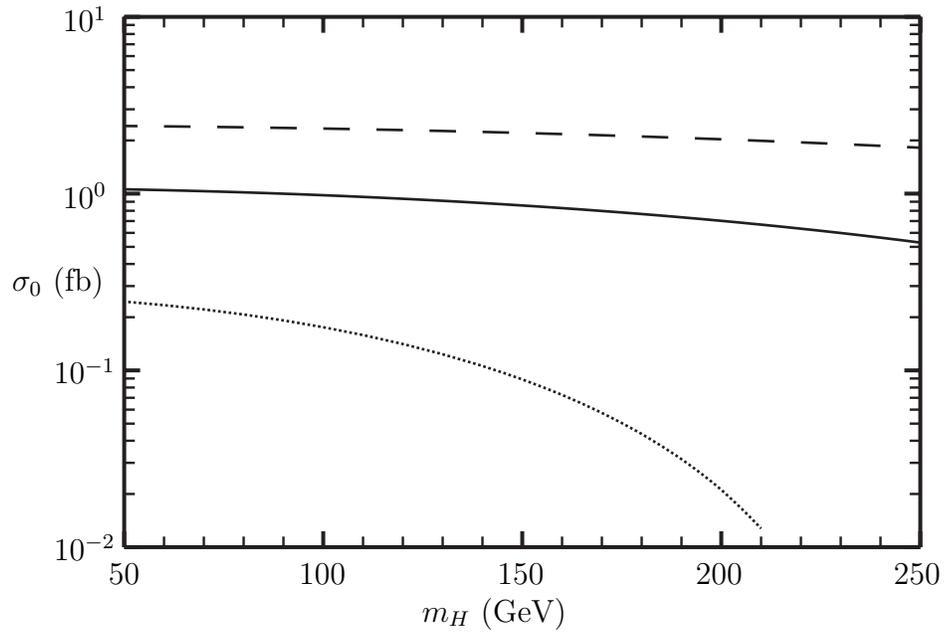
Fig. 1. The cross section $\sigma_0$ as a function of $m_H$ 
at $\sqrt{s}=0.5$ (dotted), $1.0$ (solid) and $1.5$ (dashed) TeV 
respectively.

\noindent
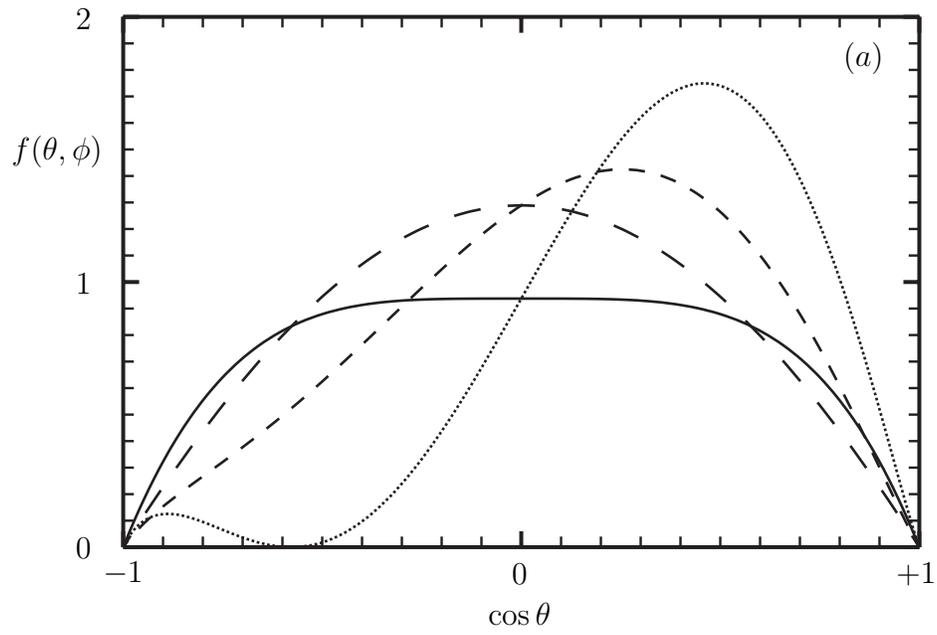
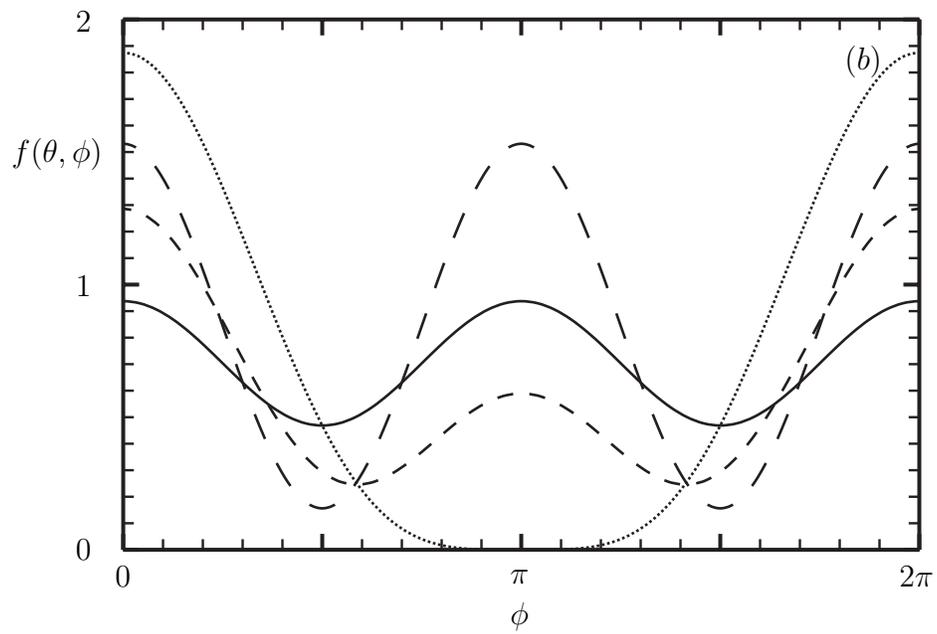
Fig. 2. The distribution $f(\theta,\phi)$ as a function of 
$\cos\theta$ at $\phi=\pi/4$ (panel $a$) and as a function of 
$\phi$ at $\theta=\pi/4$ (panel $b$). 
The solid, dotted and short-dashed curves
are for the parameter sets $(1)$, $(2)$ and $(3)$ respectively.
Also shown (long-dashed) in the panel $a$ ($b$) is the distribution
further averaged over $\phi$ ($\theta$) for the parameter set $(3)$.

\noindent
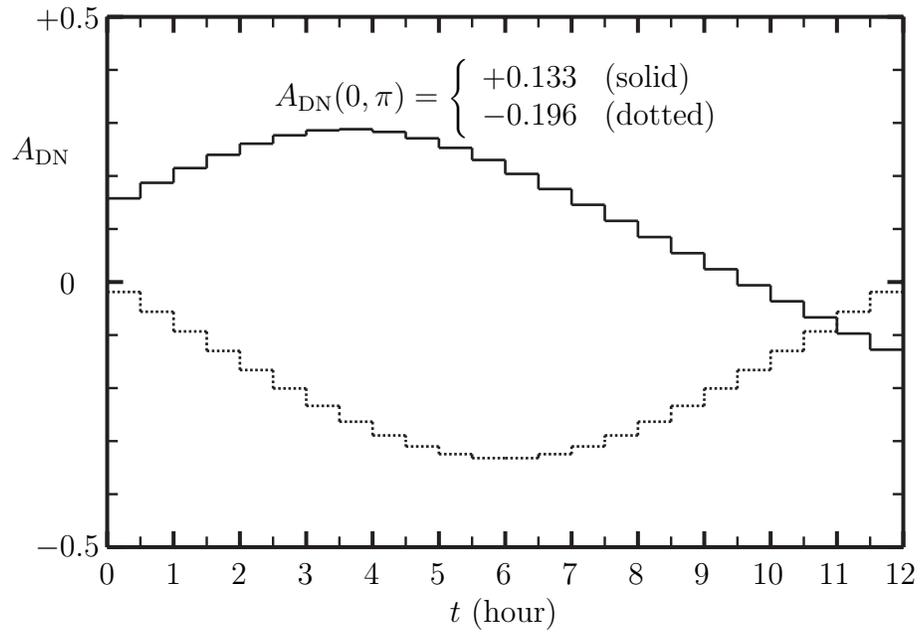
Fig. 3. Histograms of the day-night asymmetry $A_{\rm DN}$ as a 
function of time $t$. 
The solid and dotted curves
are for the parameter sets $(3)$ and $(4)$ respectively.
\newpage
\begin{center}
\begin{picture}(350,300)(0,0)

\SetOffset(40,50)\SetWidth{1.}
\LinAxis(0,0)(300,0)(4,5,5,0,1.5)
\LinAxis(0,200)(300,200)(4,5,-5,0,1.5)
\LogAxis(0,0)(0,200)(3,-5,0,1.5)
\LogAxis(300,0)(300,200)(3,5,0,1.5)
\Text(0,-10)[]{$50$}
\Text(75,-10)[]{$100$}
\Text(150,-10)[]{$150$}
\Text(225,-10)[]{$200$}
\Text(300,-10)[]{$250$}
\Text(140,-25)[]{$m_H$ (GeV)}
\Text(-15,0)[]{$10^{-2}$}
\Text(-15,66.67)[]{$10^{-1}$}
\Text(-15,133.3)[]{$10^{0}$}
\Text(-15,200)[]{$10^{1}$}
\Text(-25,100)[]{$\sigma_0$ (fb)}
\Text(140,-50)[]{Figure $1$}
\DashCurve{(0.000,92.608)(15.00,91.269)(30.00,89.653)
(45.00,87.743)(59.99,85.516)(75.00,82.943)
(90.00,79.988)(105.0,76.607)(120.0,72.744)
(135.0,68.326)(150.0,63.259)(165.0,57.421)
(180.0,50.639)(195.0,42.677)(210.0,33.183)
(225.0,21.613)(240.0,7.0529)}{1}
\Curve{(0.000,134.97)(15.00,134.65)(30.00,134.26)
(45.00,133.82)(59.99,133.31)(75.00,132.74)
(90.00,132.10)(105.0,131.40)(120.0,130.63)
(135.0,129.79)(150.0,128.87)(165.0,127.88)
(180.0,126.80)(195.0,125.65)(210.0,124.41)
(225.0,123.08)(240.0,121.65)(255.0,120.12)
(270.0,118.49)(285.0,116.74)(300.0,114.87)}
\DashCurve{(0.000,158.85)(15.00,158.71)(30.00,158.54)
(45.00,158.35)(59.99,158.12)(75.00,157.88)
(90.00,157.60)(105.0,157.30)(120.0,156.97)
(135.0,156.61)(150.0,156.22)(165.0,155.80)
(180.0,155.36)(195.0,154.88)(210.0,154.37)
(225.0,153.84)(240.0,153.26)(255.0,152.66)
(270.0,152.03)(285.0,151.36)(300.0,150.65)}{10}
\end{picture}\\
\end{center}
\newpage
\begin{center}
\begin{picture}(350,600)(0,0)
\SetOffset(10,80)\SetWidth{1.5}

\SetOffset(40,350)\SetWidth{1.}
\LinAxis(0,0)(300,0)(2,10,5,0,1.5)
\LinAxis(0,200)(300,200)(2,10,-5,0,1.5)
\LinAxis(0,200)(0,0)(2,10,5,0,1.5)
\LinAxis(300,0)(300,200)(2,10,5,0,1.5)
\Text(0,-10)[]{$-1$}
\Text(150,-10)[]{$0$}
\Text(300,-10)[]{$+1$}
\Text(150,-25)[]{$\cos\theta$}
\Text(-15,0)[]{$0$}
\Text(-15,100)[]{$1$}
\Text(-15,200)[]{$2$}
\Text(-25,150)[]{$f(\theta,\phi)$}
\Text(280,185)[]{$(a)$}
\Curve{(300.00,0.000000)(299.67,0.800330)(298.71,3.167252)
(297.11,7.000515)(294.88,12.13949)(292.03,18.37236)
(288.58,25.44825)(284.53,33.09127)(279.90,41.01586)
(274.72,48.94217)(269.00,56.61070)(262.77,63.79519)
(256.06,70.31295)(248.90,76.03210)(241.31,80.87514)
(233.33,84.81895)(225.00,87.89111)(216.34,90.16292)
(207.40,91.73976)(198.21,92.74945)(188.82,93.32953)
(179.26,93.61430)(169.57,93.72278)(159.81,93.74816)
(150.00,93.74977)(140.18,93.74796)(130.42,93.72238)
(120.73,93.61372)(111.17,93.32880)(101.78,92.74862)
(92.597,91.73885)(83.656,90.16199)(75.000,87.89018)
(66.664,84.81807)(58.686,80.87434)(51.098,76.03142)
(43.934,70.31241)(37.224,63.79478)(30.997,56.61043)
(25.279,48.94203)(20.096,41.01584)(15.469,33.09134)
(11.418,25.44839)(7.9606,18.37253)(5.1112,12.13966)
(2.8822,7.000675)(1.2833,3.167372)(0.3211,0.800394)
(0.0000,1.32E-09)}
\DashCurve{(300.00,0.000000)(299.67,0.874353)(298.71,3.751865)
(297.11,8.931550)(294.88,16.58010)(292.03,26.71160)
(288.58,39.17746)(284.53,53.66723)(279.90,69.72054)
(274.72,86.74946)(269.00,104.0699)(262.77,120.9401)
(256.06,136.6037)(248.90,150.3344)(241.31,161.4799)
(233.33,169.5011)(225.00,174.0056)(216.34,174.7712)
(207.40,171.7602)(198.21,165.1211)(188.82,155.1802)
(179.26,142.4205)(169.57,127.4534)(159.81,110.9801)
(150.00,93.75035)(140.18,76.51717)(130.42,59.99286)
(120.73,44.80843)(111.17,31.47899)(101.78,20.37759)
(92.597,11.71891)(83.656,5.553977)(75.000,1.775813)
(66.664,0.135820)(58.686,0.269351)(51.098,1.728712)
(43.934,4.021180)(37.224,6.649309)(30.997,9.150670)
(25.279,11.13421)(20.096,12.31067)(15.469,12.51496)
(11.418,11.71882)(7.9606,10.03303)(5.1112,7.698869)
(2.8822,5.069530)(1.2833,2.582707)(0.3211,0.726359)
(0.0000,1.32E-09)}{1}
\DashCurve{(300.00,0.000000)(299.67,0.528989)(298.71,2.217096)
(297.11,5.194458)(294.88,9.552156)(292.03,15.33097)
(288.58,22.51284)(284.53,31.01563)(279.90,40.69180)
(274.72,51.33104)(269.00,62.66694)(262.77,74.38728)
(256.06,86.14743)(248.90,97.58608)(241.31,108.3423)
(233.33,118.0731)(225.00,126.4699)(216.34,133.2738)
(207.40,138.2874)(198.21,141.3842)(188.82,142.5133)
(179.26,141.7002)(169.57,139.0431)(159.81,134.7057)
(150.00,128.9064)(140.18,121.9045)(130.42,113.9849)
(120.73,105.4421)(111.17,96.56448)(101.78,87.61913)
(92.597,78.83997)(83.656,70.41796)(75.000,62.49514)
(66.664,55.16234)(58.686,48.46065)(51.098,42.38646)
(43.934,36.89964)(37.224,31.93396)(30.997,27.40922)
(25.279,23.24380)(20.096,19.36695)(15.469,15.72967)
(11.418,12.31339)(7.9606,9.135777)(5.1112,6.253279)
(2.8822,3.759955)(1.2833,1.782844)(0.3211,0.474032)
(0.0000,8.25E-10)}{6}
\DashCurve{(300.00,0.000000)(299.67,0.501492)(298.71,1.999936)
(297.11,4.477161)(294.88,7.902662)(292.03,12.23331)
(288.58,17.41304)(284.53,23.37257)(279.90,30.02929)
(274.72,37.28732)(269.00,45.03797)(262.77,53.16050)
(256.06,61.52338)(248.90,69.98610)(241.31,78.40130)
(233.33,86.61752)(225.00,94.48230)(216.34,101.8456)
(207.40,108.5634)(198.21,114.5013)(188.82,119.5386)
(179.26,123.5708)(169.57,126.5136)(159.81,128.3048)
(150.00,128.9061)(140.18,128.3048)(130.42,126.5137)
(120.73,123.5710)(111.17,119.5388)(101.78,114.5016)
(92.597,108.5636)(83.656,101.8459)(75.000,94.48260)
(66.664,86.61784)(58.686,78.40163)(51.098,69.98644)
(43.934,61.52372)(37.224,53.16083)(30.997,45.03829)
(25.279,37.28763)(20.096,30.02957)(15.469,23.37283)
(11.418,17.41327)(7.9606,12.23350)(5.1112,7.902820)
(2.8822,4.477281)(1.2833,2.000017)(0.3211,0.501533)
(0.0000,8.25E-10)}{10}

\SetOffset(40,50)\SetWidth{1.}
\LinAxis(0,0)(300,0)(4,5,5,0,1.5)
\LinAxis(0,200)(300,200)(4,5,-5,0,1.5)
\LinAxis(0,200)(0,0)(2,10,5,0,1.5)
\LinAxis(300,0)(300,200)(2,10,5,0,1.5)
\Text(0,-10)[]{$0$}
\Text(150,-10)[]{$\pi$}
\Text(300,-10)[]{$2\pi$}
\Text(150,-25)[]{$\phi$}
\Text(-15,0)[]{$0$}
\Text(-15,100)[]{$1$}
\Text(-15,200)[]{$2$}
\Text(-25,150)[]{$f(\theta,\phi)$}
\Text(280,185)[]{$(b)$}
\Text(140,-50)[]{Figure $2$}
\Curve{(0.0000,93.750)(6.2500,92.951)(12.500,90.610)
(18.750,86.885)(25.000,82.031)(31.250,76.379)
(37.500,70.312)(43.750,64.246)(50.000,58.594)
(56.250,53.740)(62.500,50.015)(68.750,47.673)
(75.000,46.875)(81.250,47.673)(87.500,50.015)
(93.750,53.739)(100.00,58.593)(106.25,64.246)
(112.50,70.312)(118.75,76.378)(125.00,82.030)
(131.25,86.884)(137.50,90.609)(143.75,92.950)
(150.00,93.749)(156.25,92.950)(162.50,90.609)
(168.75,86.884)(175.00,82.030)(181.25,76.378)
(187.50,70.312)(193.75,64.246)(200.00,58.593)
(206.25,53.739)(212.50,50.015)(218.75,47.673)
(225.00,46.875)(231.25,47.673)(237.50,50.015)
(243.75,53.739)(250.00,58.593)(256.25,64.246)
(262.50,70.312)(268.75,76.378)(275.00,82.031)
(281.25,86.885)(287.50,90.610)(293.75,92.951)
(300.00,93.750)}
\DashCurve{(0.0000,187.49)(6.2500,185.89)(12.500,181.16)
(18.750,173.49)(25.000,163.22)(31.250,150.75)
(37.500,136.60)(43.750,121.31)(50.000,105.46)
(56.250,89.616)(62.500,74.279)(68.750,59.910)
(75.000,46.875)(81.250,35.436)(87.500,25.750)
(93.750,17.863)(100.00,11.718)(106.25,7.1751)
(112.50,4.0213)(118.75,2.0017)(125.00,0.8414)
(131.25,0.2716)(137.50,5.4434E-02)(143.75,3.4331E-03)
(150.00,3.2997E-10)(156.25,3.4326E-03)(162.50,5.4430E-02)
(168.75,0.2716)(175.00,0.8413)(181.25,2.0017)
(187.50,4.0212)(193.75,7.1750)(200.00,11.718)
(206.25,17.863)(212.50,25.750)(218.75,35.436)
(225.00,46.874)(231.25,59.910)(237.50,74.278)
(243.75,89.615)(250.00,105.46)(256.25,121.31)
(262.50,136.60)(268.75,150.75)(275.00,163.22)
(281.25,173.49)(287.50,181.16)(293.75,185.89)
(300.00,187.49)}{1}
\DashCurve{(0.0000,128.57)(6.2500,127.17)(12.500,123.06)
(18.750,116.48)(25.000,107.79)(31.250,97.491)
(37.500,86.147)(43.750,74.381)(50.000,62.822)
(56.250,52.062)(62.500,42.627)(68.750,34.940)
(75.000,29.296)(81.250,25.849)(87.500,24.601)
(93.750,25.409)(100.00,27.998)(106.25,31.983)
(112.50,36.899)(118.75,42.236)(125.00,47.478)
(131.25,52.138)(137.50,55.795)(143.75,58.126)
(150.00,58.926)(156.25,58.126)(162.50,55.795)
(168.75,52.138)(175.00,47.478)(181.25,42.236)
(187.50,36.899)(193.75,31.983)(200.00,27.998)
(206.25,25.409)(212.50,24.601)(218.75,25.849)
(225.00,29.296)(231.25,34.940)(237.50,42.627)
(243.75,52.061)(250.00,62.821)(256.25,74.381)
(262.50,86.146)(268.75,97.491)(275.00,107.79)
(281.25,116.48)(287.50,123.06)(293.75,127.17)
(300.00,128.57)}{6}
\DashCurve{(0.0000,153.124)(6.2500,150.782)(12.500,143.914)
(18.750,132.988)(25.000,118.750)(31.250,102.168)
(37.500,84.3750)(43.750,66.5812)(50.000,50.0001)
(56.250,35.7615)(62.500,24.8358)(68.750,17.9676)
(75.000,15.6250)(81.250,17.9675)(87.500,24.8356)
(93.750,35.7612)(100.00,49.9997)(106.25,66.5809)
(112.50,84.3746)(118.75,102.168)(125.00,118.749)
(131.25,132.988)(137.50,143.914)(143.75,150.782)
(150.00,153.124)(156.25,150.782)(162.50,143.914)
(168.75,132.988)(175.00,118.750)(181.25,102.169)
(187.50,84.3754)(193.75,66.5816)(200.00,50.0004)
(206.25,35.7617)(212.50,24.8360)(218.75,17.9677)
(225.00,15.6250)(231.25,17.9674)(237.50,24.8354)
(243.75,35.7610)(250.00,49.9994)(256.25,66.5805)
(262.50,84.3743)(268.75,102.168)(275.00,118.749)
(281.25,132.988)(287.50,143.913)(293.75,150.782)
(300.00,153.124)}{10}
\end{picture}\\
\end{center}
\newpage
\begin{center}
\begin{picture}(350,300)(0,0)
\SetOffset(10,80)\SetWidth{1.5}

\SetOffset(40,50)\SetWidth{1.}
\LinAxis(0,0)(300,0)(12,2,5,0,1.5)
\LinAxis(0,200)(300,200)(12,2,-5,0,1.5)
\LinAxis(0,200)(0,0)(2,5,5,0,1.5)
\LinAxis(300,0)(300,200)(2,5,5,0,1.5)
\Text(0,-10)[]{$0$}\Text(25,-10)[]{$1$}\Text(50,-10)[]{$2$}
\Text(75,-10)[]{$3$}\Text(100,-10)[]{$4$}\Text(125,-10)[]{$5$}
\Text(150,-10)[]{$6$}\Text(175,-10)[]{$7$}\Text(200,-10)[]{$8$}
\Text(225,-10)[]{$9$}\Text(250,-10)[]{$10$}\Text(275,-10)[]{$11$}
\Text(300,-10)[]{$12$}
\Text(150,-25)[]{$t$ (hour)}
\Text(-15,0)[]{$-0.5$}
\Text(-15,100)[]{$0$}
\Text(-15,200)[]{$+0.5$}
\Text(-25,150)[]{$A_{\rm DN}$}
\Text(150,170)[]{$A_{\rm DN}(0,\pi)=\left\{
\begin{array}{ll}
+0.133&({\rm solid})\\
-0.196&({\rm dotted})
\end{array}\right.$}
\Text(140,-50)[]{Figure $3$}
\Line(0.000,131.55)(12.50,131.55)\Line(12.50,131.55)(12.50,137.42)
\Line(12.50,137.42)(25.00,137.42)\Line(25.00,137.42)(25.00,142.97)
\Line(25.00,142.97)(37.50,142.97)\Line(37.50,142.97)(37.50,147.97)
\Line(37.50,147.97)(50.00,147.97)\Line(50.00,147.97)(50.00,152.18)
\Line(50.00,152.18)(62.50,152.18)\Line(62.50,152.18)(62.50,155.33)
\Line(62.50,155.33)(75.00,155.33)\Line(75.00,155.33)(75.00,157.20)
\Line(75.00,157.20)(87.50,157.20)\Line(87.50,157.20)(87.50,157.64)
\Line(87.50,157.64)(100.0,157.64)\Line(100.0,157.64)(100.0,156.61)
\Line(100.0,156.61)(112.5,156.61)\Line(112.5,156.61)(112.5,154.20)
\Line(112.5,154.20)(125.0,154.20)\Line(125.0,154.20)(125.0,150.59)
\Line(125.0,150.59)(137.5,150.59)\Line(137.5,150.59)(137.5,146.03)
\Line(137.5,146.03)(150.0,146.03)\Line(150.0,146.03)(150.0,140.78)
\Line(150.0,140.78)(162.5,140.78)\Line(162.5,140.78)(162.5,135.08)
\Line(162.5,135.08)(175.0,135.08)\Line(175.0,135.08)(175.0,129.13)
\Line(175.0,129.13)(187.5,129.13)\Line(187.5,129.13)(187.5,123.06)
\Line(187.5,123.06)(200.0,123.06)\Line(200.0,123.06)(200.0,116.96)
\Line(200.0,116.96)(212.5,116.96)\Line(212.5,116.96)(212.5,110.88)
\Line(212.5,110.88)(225.0,110.88)\Line(225.0,110.88)(225.0,104.82)
\Line(225.0,104.82)(237.5,104.82)\Line(237.5,104.82)(237.5,98.779)
\Line(237.5,98.779)(250.0,98.779)\Line(250.0,98.779)(250.0,92.730)
\Line(250.0,92.730)(262.5,92.730)\Line(262.5,92.730)(262.5,86.661)
\Line(262.5,86.661)(275.0,86.661)\Line(275.0,86.661)(275.0,80.569)
\Line(275.0,80.569)(287.5,80.569)\Line(287.5,80.569)(287.5,74.476)
\Line(287.5,74.476)(300.0,74.476)
\DashLine(0.000,96.262)(12.50,96.262){1}\DashLine(12.50,96.262)(12.50,88.796){1}
\DashLine(12.50,88.796)(25.00,88.796){1}\DashLine(25.00,88.796)(25.00,81.367){1}
\DashLine(25.00,81.367)(37.50,81.367){1}\DashLine(37.50,81.367)(37.50,74.015){1}
\DashLine(37.50,74.015)(50.00,74.015){1}\DashLine(50.00,74.015)(50.00,66.809){1}
\DashLine(50.00,66.809)(62.50,66.809){1}\DashLine(62.50,66.809)(62.50,59.855){1}
\DashLine(62.50,59.855)(75.00,59.855){1}\DashLine(75.00,59.855)(75.00,53.298){1}
\DashLine(75.00,53.298)(87.50,53.298){1}\DashLine(87.50,53.298)(87.50,47.322){1}
\DashLine(87.50,47.322)(100.0,47.322){1}\DashLine(100.0,47.322)(100.0,42.145){1}
\DashLine(100.0,42.145)(112.5,42.145){1}\DashLine(112.5,42.145)(112.5,37.995){1}
\DashLine(112.5,37.995)(125.0,37.995){1}\DashLine(125.0,37.995)(125.0,35.087){1}
\DashLine(125.0,35.087)(137.5,35.087){1}\DashLine(137.5,35.087)(137.5,33.586){1}
\DashLine(137.5,33.586)(150.0,33.586){1}
\DashLine(150.0,33.586)(162.5,33.586){1}\DashLine(162.5,33.586)(162.5,35.086){1}
\DashLine(162.5,35.086)(175.0,35.086){1}\DashLine(175.0,35.086)(175.0,37.995){1}
\DashLine(175.0,37.995)(187.5,37.995){1}\DashLine(187.5,37.995)(187.5,42.145){1}
\DashLine(187.5,42.145)(200.0,42.145){1}\DashLine(200.0,42.145)(200.0,47.322){1}
\DashLine(200.0,47.322)(212.5,47.322){1}\DashLine(212.5,47.322)(212.5,53.298){1}
\DashLine(212.5,53.298)(225.0,53.298){1}\DashLine(225.0,53.298)(225.0,59.855){1}
\DashLine(225.0,59.855)(237.5,59.855){1}\DashLine(237.5,59.855)(237.5,66.809){1}
\DashLine(237.5,66.809)(250.0,66.809){1}\DashLine(250.0,66.809)(250.0,74.014){1}
\DashLine(250.0,74.014)(262.5,74.014){1}\DashLine(262.5,74.014)(262.5,81.366){1}
\DashLine(262.5,81.366)(275.0,81.366){1}\DashLine(275.0,81.366)(275.0,88.796){1}
\DashLine(275.0,88.796)(287.5,88.796){1}\DashLine(287.5,88.796)(287.5,96.262){1}
\DashLine(287.5,96.262)(300.0,96.262){1}
\end{picture}\\
\end{center}

\end{document}